\documentclass[aps,twocolumn,showpacs,preprintnumbers,floats]{revtex4}\def\@cite#1#2{\textsuperscript{[{#1\if@tempswa , #2\fi}]}}
\usepackage{mathrsfs}

\usepackage{longtable,lscape}
\usepackage{savesym}
\usepackage{txfonts}
\savesymbol{iint}
\savesymbol{iiint}
\savesymbol{iiiint}
\savesymbol{idotsint}
\usepackage{amssymb}
\usepackage{amsmath}
\restoresymbol{TXF}{iint}
\restoresymbol{TXF}{iiint}
\restoresymbol{TXF}{iiiint}
\restoresymbol{TXF}{idotsint}
\usepackage{indentfirst}
\usepackage{graphicx,booktabs}
\usepackage{braket}
\usepackage{multirow}
\usepackage{color}
\usepackage{subfigure}
\usepackage{float}
\usepackage{bm}
\usepackage{epsfig}
\usepackage[colorlinks, citecolor=blue,anchorcolor=red,menucolor=red, linkcolor=red,filecolor=red,urlcolor=blue,frenchlinks=red]{hyperref}

\begin{document}

	\title{Strong decays of low-lying  $D$-wave $\Xi_b/\Xi_b'$ baryons with QPC model  }
	\author{Yu-Hui Zhou$^{1}$, Wen-Jia Wang$^{1}$, Li-Ye Xiao$^{1}$~\footnote {E-mail: lyxiao@ustb.edu.cn}, Xian-Hui Zhong$^{2,3}$~\footnote {E-mail: zhongxh@hunnu.edu.cn}}
\affiliation{ 1)Institute of Theoretical Physics, University of Science and Technology Beijing,
Beijing 100083, China}
\affiliation{ 2) Department of
Physics, Hunan Normal University, and Key Laboratory of
Low-Dimensional Quantum Structures and Quantum Control of Ministry
of Education, Changsha 410081, China }
\affiliation{ 3) Synergetic
Innovation Center for Quantum Effects and Applications (SICQEA),
Hunan Normal University, Changsha 410081, China}
		
\begin{abstract}
For further decoding the inner structure of the two excited $\Xi_b$ states observed by LHCb, we perform a systematical study of the strong decays of the low-lying $1D$-wave $\Xi_b$ and $\Xi_b'$ excitations using the quark pair creation model within the $j-j$ coupling scheme. Combining with the measured masses and decay properties of $\Xi_{b}(6327)^{0}$ and $\Xi_{b}(6327)^{0}$, the two excited states can be explained as $1D$ $\lambda$-mode $\Xi_b$ states $\Xi_{b}\ket{J^{P}=\frac{3}{2}^{+},2}_{\lambda\lambda}$ and $\Xi_{b}\ket{J^{P}=\frac{5}{2}^{+},2}_{\lambda\lambda}$, respectively. If such a view were correct, $\Xi_b'\pi$ and $\Xi_b'^*\pi$ could be another interesting channels for experimental exploring of the $\Xi_{b}(6327)^{0}$ and $\Xi_{b}(6327)^{0}$, respectively. Those calculations are good consistent with the results within the chiral quark model. In addition, for the other missing $1D$-wave $\Xi_b$ and $\Xi_b'$ excitations, our predictions indicate that:(i) the two $\rho$-mode $1D$ $\Xi_b$ states are likely to be moderate states with a width of $\Gamma\sim50$ MeV. The $J^P=3/2^+$ state dominantly decays into $\Sigma_bK$ and $\Xi_b'\pi$, while the $J^P=5/2^+$ state decays primarily through $\Sigma_b^*K$ and $\Xi_b'^*\pi$. (ii) The $\lambda$-mode $1D$ $\Xi_b'$ states may be moderate states with a widths of about several to dozens of MeV. Most of the $\lambda$-mode $1D$ $\Xi_b'$ states mainly decay into the $1P$-wave bottomed baryon via the pionic decay processes. Meanwhile, several $\lambda$-mode $1D$ $\Xi_b'$ states have significant decay rates into $\Lambda B$. (iii) While, the $\rho$-mode $1D$ $\Xi_b'$ states are predicted to be very broad states with a width of about several hundreds MeV. It will be a great challenge to explore the $\rho$-mode $1D$ $\Xi_b'$ states in experiments for their broad widths.

	\end{abstract}
		\maketitle

\section{Introduction}
Based on the constituent quark model~\cite{Gell-Mann:1964ewy,Zweig:1964ruk}, singly bottom baryon contains a $b$ quark and two light quarks. If the two light quarks are $u$ or $d$ quarks, these singly bottom baryons are classified into the $\Lambda_b$ and $\Sigma_b$ baryon families with the light diquark spin $s_{\rho}$ being $0$ and $1$, respectively. When the two light quarks both are $s$ quarks, then these singly bottom baryons form $\Omega_b$ family with the light diquark spin $s_{\rho}$ being $1$. A singly bottom baryon containing one $s$ quark forms the $\Xi_b$ or $\Xi_b'$ family according to whether the light diquark spin $s_{\rho}$ is $0$ or $1$. The research of singly bottom baryon gets much attention for the mass of bottom quark $m_b$ being marginally larger than the QCD energy scale. Hence, studies of their various properties can provide valuable validation to the understanding of the non-perturbative energy region quantum chromodynamics, which features the behavior of strong interactions.

During the past few decades, experimenters have made important progress in establishing and perfecting the singly bottom baryon spectrum. Until now, all the $1S$ singly bottom baryons are well established in experiments except the $\Omega_b^*$~\cite{ParticleDataGroup:2020ssz}. Meanwhile, there are several higher candidates observed by experiments: nine $1P$ candidates, one $2S$ candidate and two $1D$ candidates. These observed singly bottom baryons may provide an access  point for study of quantum chromodynamics and have attracted significant attention from the hadron physics community. Their mass spectra and strong decay properties were extensively explored with various theoretical methods and models ~\cite{Ebert:2011kk,Roberts:2007ni,Yoshida:2015tia,Liu:2007fg,Chen:2016phw,Mao:2017wbz,Cui:2019dzj,Yang:2019cvw,Mao:2020jln,Shifman:1978bx,
Reinders:1984sr,Balitsky:1989ry,Braun:1988qv,Chernyak:1990ag,Ball:1998ff,Ball:2004rg,Ball:2007zt,Falk:1990yz,Yang:2020zrh,Liang:2020hbo}. More detail information can refers to reviews~\cite{Chen:2022asf,Chen:2016spr}.

%Single  bottom
%baryon is always an interesting research object  because that the mass of bottom quark $m_b$ is marginally larger than the QCD energy scale.
%During the past several years, both great experimental and theoretical progress have been made in excited single bottom baryons.
%Since  two candidates for the P-wave
%bottom baryons, $\Xi_{b}(6227)^{-}$
% and $\Sigma_{b}(6097)^{\pm}$, were
%discovered by the LHCb,
%the construction of the highly excited bottom baryon spectroscopy has been unveiled.
%As a member of single bottom baryons, $\Xi_{b}$  is active.
%$\Xi_b$ baryons can be divided into $\Xi_b$ with antisymmetric flavor configuration $\overline{3}_F$ and $\Xi_b^{'}$ with symmetric flavor configuration $6_F$.
%They are distinguished by adding a superscript to the heavier.
%It's difficult to distinguish them because of the same isopin I=1/2.
%

Last year, the LHCb collaboration reported  two narrow resonances, $\Xi_b(6327)^{0}$ and $\Xi_b(6333)^{0}$, in the $\Lambda_b^0$$K^-$$\pi^+$ mass spectrum~\cite{LHCb:2021ssn}. Their masses and decay widths are respectively
\begin{equation}
	m[\Xi_b(6327)^0]=6327.28_{-0.21}^{+0.23}\pm0.12\pm0.24\ \rm{MeV},
\end{equation}
\begin{equation}
	\Gamma[(\Xi_b(6327)^0)]=0.93_{-0.60}^{+0.74}\ \rm{MeV},
\end{equation}
\begin{equation}
	m[\Xi_b(6333)^0]=6332.69_{-0.18}^{+0.17}\pm0.03\pm0.22\ \rm{MeV},
\end{equation}
\begin{equation}
	\Gamma[(\Xi_b(6333)^0)]=0.25_{-0.25}^{+0.58}\ \rm{MeV}.
\end{equation}
Moreover, it is found that the lighter state $\Xi_b(6327)^0$ predominantly decays to $\Sigma_b^+K^-$, while the heavier state $\Xi_b(6333)^0$ has a significant decay rate into $\Sigma_b^{*+}K^-$. According to the previous study of the $\Xi_b^{(')}$ baryon spectrum~\cite{Ebert:2007nw,Ebert:2011kk,Kakadiya:2022zvy,Thakkar:2016dna,Wei:2016jyk,Chen:2019ywy,Yu:2021zvl,Yang:2020zrh,Cui:2019dzj,Jia:2019bkr,
Narodetskii:2008pn,Valcarce:2008dr,Chen:2018orb,Li:2022xtj,Yang:2022oog,Garcia-Tecocoatzi:2023btk}, the two newly observed $\Xi_b^0$ states~\cite{LHCb:2021ssn} may be $\lambda$-mode $1D$ $\Xi_b$ states(see table~\ref{table1}). In addition, the few works on strong decay properties also supported the two newly states as $\lambda$-mode $1D$ $\Xi_b$ states with $J^p=3/2^+$ and $J^P=5/2^+$~\cite{Chen:2019ywy,Yao:2018jmc,Garcia-Tecocoatzi:2023btk,Bijker:2020tns,Yu:2021zvl,Wang:2022zcy}.

In our previous work~\cite{Wang:2022zcy}, we systematically investigated the strong transitions of the low-lying $1D$-wave $\Xi_{b}/\Xi_{b}'$ resonances with chiral quark model and given an explanation of the two newly observed $\Xi_b$ states~\cite{LHCb:2021ssn}. While, we notice that for the low-lying $1D$-wave $\Xi_{b}/\Xi_{b}'$ resonances, their masses are large enough to allow the decay channels containing a heavy-light flavor meson. Hence, it is suitable to apply the quark pair creation(QPC) strong decay model to discuss those decay channels, which haven't been studied systematically. Meanwhile, for further understanding the inner structures of the $1D$-wave $\Xi_{b}/\Xi_{b}'$ baryons, it is necessary to make a comparison of the theoretical predictions with QPC model to the results with the chiral quark model~\cite{Wang:2022zcy}. Thus, in this work, we conduct a systematic analysis of the two body strong decays of the $1D$-wave $\Xi_{b}/\Xi_{b}'$ baryons with the QPC model under the $j-j$ coupling scheme, which may provide more information of their strong decays. The quark model classification, predicted masses,
and OZI-allowed decay modes  are summarized in
Table~\ref{table1}.

 This paper is organized as follows. In Sec.~\ref{model}, we briefly introduce the mechanism of QPC model and present the parameter values. Then we give our theoretical results and discussions in Sec.~\ref{results}. Finally, we give a summary of our results
 in Sec.~\ref{summary}.

 \begin{table*}
	\caption{Predicted mass of 1D-wave excited $\Xi_b$ an $\Xi^{'}_b$ baryons  in various quark models}
	
	\label{table1}
	\begin{tabular}{cccccccccccccc}
		
		\toprule[1.2pt]
		\multirow{1}{*}{Notation}& \multicolumn{6}{c}{$\rm{Quantum}$ $\rm{Number}$}& \multicolumn{6}{c}{$\rm{Mass}$}  & \multicolumn{1}{c}{$\rm{Decay}$ $\rm{channel}$}  \\
		\hline
		$\Xi_b$${\ket{J^P,j}}_{\lambda(\rho)}$&$l_{\lambda}$&$l_{\rho}$&$L$&$s_{\rho}$&$j$&$J^P$&GIM~\cite{Li:2022xtj}&hCQM~\cite{Thakkar:2016dna}& QPM~\cite{Chen:2019ywy}& RQM~\cite{Ebert:2011kk}&hCQM~\cite{Kakadiya:2022zvy}&QCD~\cite{Yu:2021zvl}& \\
		
		\hline
		$\Xi_b$${\ket{J^P=\frac{3}{2}^+,2}}_{\lambda\lambda}$&2&0&2&0&2&$\frac{3}{2}^{+}$
		&6320&6386&6327&6366&6243&6340 &$\Xi_b^{'(*)}\pi,\Sigma_b^{(*)}K$\\
		$\Xi_b$${\ket{J^P=\frac{5}{2}^+,2}}_{\lambda\lambda}$$$  &2&0&2&0&2&$\frac{5}{2}^{+}$	&6327&6369&6330&6373&6240&6360  \\

		$\Xi_b$${\ket{J^P=\frac{3}{2}^+,2}}_{\rho\rho}$&0&2&2&0&2&$\frac{3}{2}^{+}$  &&& &&&6420&\\
		
		$\Xi_b$${\ket{J^P=\frac{5}{2}^+,2}}_{\rho\rho}$	&0&2&2&0&2&$\frac{5}{2}^{+}$
		&&&& &&6430  \\
		\hline
		$\Xi_b^{'}$${\ket{J^P=\frac{1}{2}^+,1}}_{\lambda\lambda}$&2&0&2&1&1&$\frac{1}{2}^{+}$
		&6460 &&6486&6447&6380&&$\Xi_b\pi,\Xi_b^{'(*)}\pi,\Xi_b^{'(*)}\eta,\Sigma_b^{(*)}K,\Lambda_bK,\Lambda B,\Sigma B,$\\

		$\Xi_b^{'}$${\ket{J^P=\frac{3}{2}^+,1}}_{\lambda\lambda}$&2&0&2&1&1&$\frac{3}{2}^{+}$  	
		&6466 &&6488&6459&6375&&$\ket{\Lambda_b \ P_{\lambda}}K,\ket{\Xi_b(\Xi^{'}_{b})\ P_{\lambda}}\pi$\\
		$\Xi_b^{'}$${\ket{J^P=\frac{3}{2}^+,2}}_{\lambda\lambda}$&2&0&2&1&2&$\frac{3}{2}^{+}$  	
		&6460 &&6456&6431&6377&& \\
		
		$\Xi_b^{'}$${\ket{J^P=\frac{5}{2}^+,2}}_{\lambda\lambda}$	&2&0&2&1&2&$\frac{5}{2}^{+}$  	
		&6466&&6457&6432&6371&&\\
		$\Xi_b^{'}$${\ket{J^P=\frac{5}{2}^+,3}}_{\lambda\lambda}$	&2&0&2&1&3&$\frac{5}{2}^{+}$  	
		&6460 &&6407&6420&6373 \\
		
		$\Xi_b^{'}$${\ket{J^P=\frac{7}{2}^+,3}}_{\lambda\lambda}$	&2&0&2&1&3&$\frac{7}{2}^{+}$  	
		&6467 &&6408&6414&6368  \\
		\hline

		$\Xi_b^{'}$${\ket{J^P=\frac{1}{2}^+,1}}_{\rho\rho}$&0&2&2&1&1&$\frac{1}{2}^{+}$ &&&&&&& $\Xi_b\pi,\Xi_b^{'(*)}\pi,\Xi_b^{'(*)}\eta,\Sigma_b^{(*)}K,\Lambda_bK,$\\
		$\Xi_b^{'}$${\ket{J^P=\frac{3}{2}^+,1}}_{\rho\rho}$&0&2&2&1&1&$\frac{3}{2}^{+}$ &&&&&&&$\ket{\Lambda_b \ P_{\lambda}}K,\ket{\Xi_b(\Xi^{'}_{b})\ P_{\lambda}}\pi$\\
		$\Xi_b^{'}$${\ket{J^P=\frac{3}{2}^+,2}}_{\rho\rho}$&0&2&2&1&2&$\frac{3}{2}^{+}$ &&&&&&&\\
		$\Xi_b^{'}$${\ket{J^P=\frac{5}{2}^+,2}}_{\rho\rho}$&0&2&2&1&2&$\frac{5}{2}^{+}$ &&&&&&&\\
		$\Xi_b^{'}$${\ket{J^P=\frac{5}{2}^+,3}}_{\rho\rho}$&0&2&2&1&3&$\frac{5}{2}^{+}$ &&&&&&&\\
		$\Xi_b^{'}$${\ket{J^P=\frac{7}{2}^+,3}}_{\rho\rho}$&0&2&2&1&3&$\frac{7}{2}^{+}$ &&&&&&&\\
		\bottomrule[1.2pt]
	\end{tabular}
\end{table*}

\section{QPC model}\label{model}
The QPC model~\cite{Micu:1968mk,R1970Regge}, also famous as $^3P_0$ model, plays an important role in studying the strong decay behaviors of the mesons and baryons~\cite{Chen:2007xf,Zhao:2016qmh,Chen:2017aqm,Chen:2017gnu,Chen:2018orb,Limphirat:2010zz,He:2021xrh,Lu:2018utx,Xiao:2017dly,Weng:2018ebv,
He:2021iwx,Liang:2020hbo,Lu:2019rtg,Liang:2019aag,Ye:2017yvl,Yang:2018lzg}.
The main idea of this model is that strong decays take place via the creation of quark-antiquark pair from the vacuum with quantum number $0^{++}$.
For baryon decays, one quark of the initial baryon A regroups with the created antiquark to form a meson C, and the other two quarks regroup with
the created quark to form a daughter baryon B.
\begin{figure}[h]
	\begin{center}
		\subfigure[]{\begin{minipage}{0.32\linewidth}
				\centering
				\includegraphics[width=1in,height=1in]{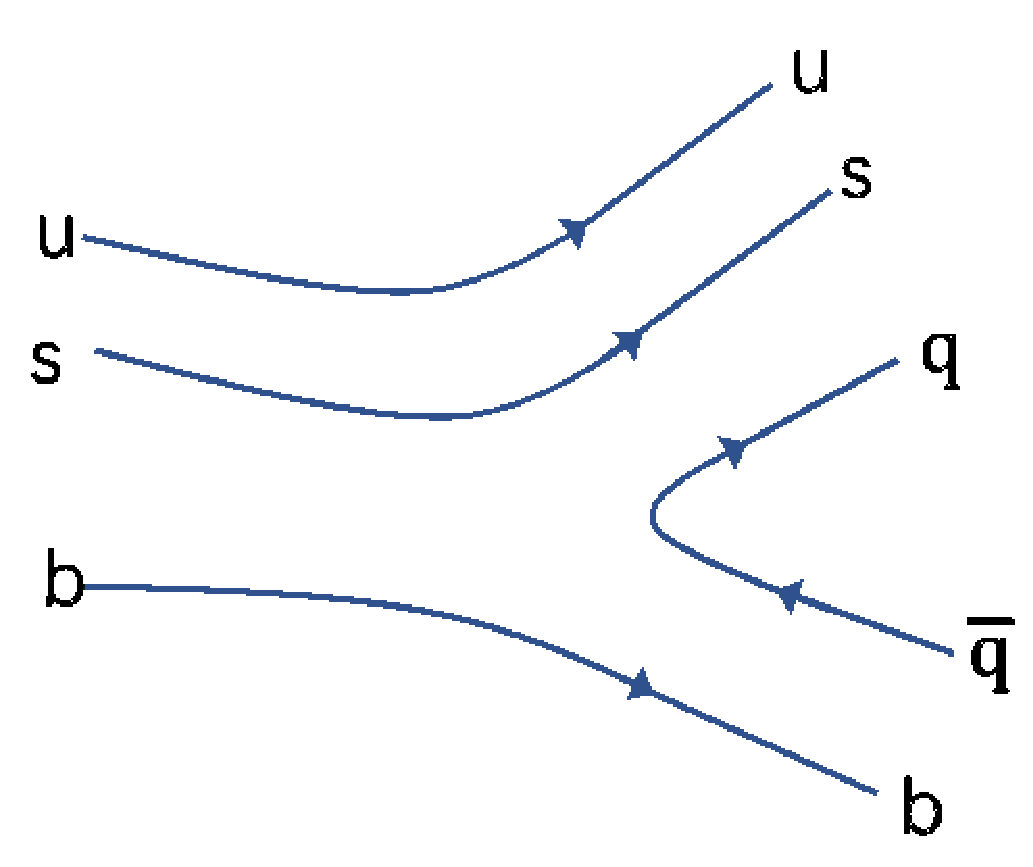}
		\end{minipage}}
		\subfigure[]{\begin{minipage}{0.32\linewidth}
				\centering
				\includegraphics[width=1in,height=1in]{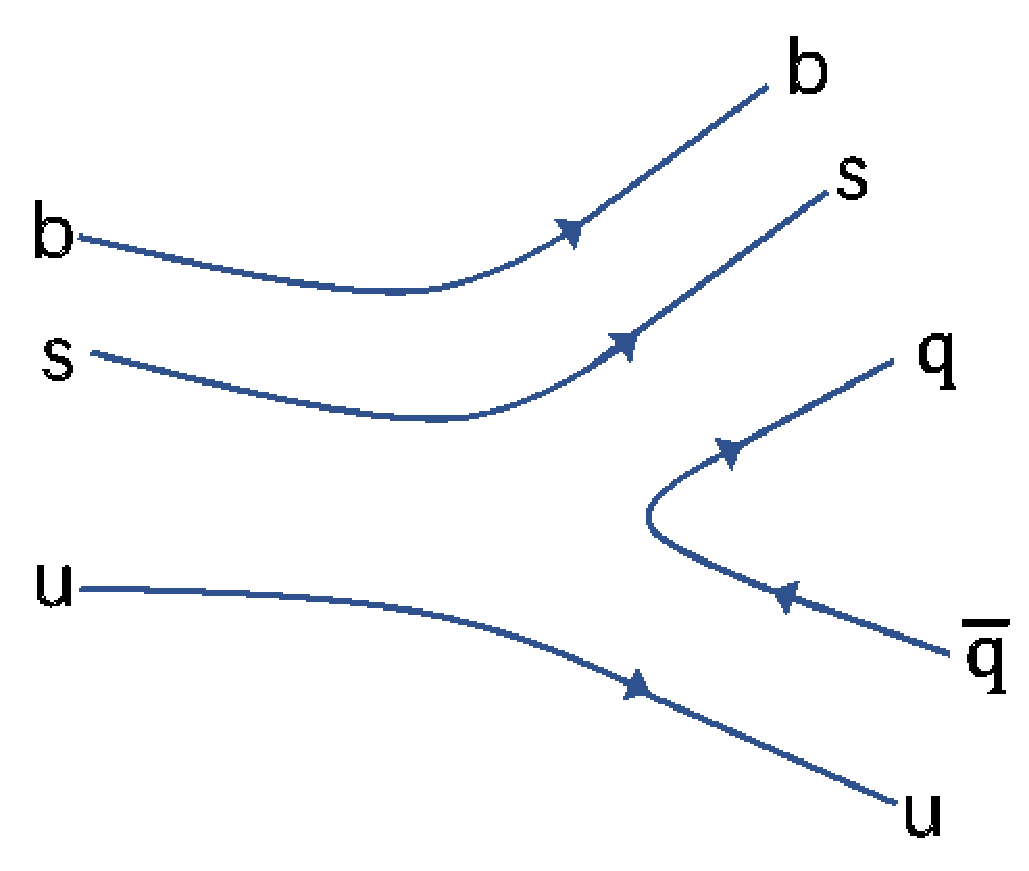}
		\end{minipage}}
		\subfigure[]{\begin{minipage}{0.32\linewidth}
				\centering
				\includegraphics[width=1in,height=1in]{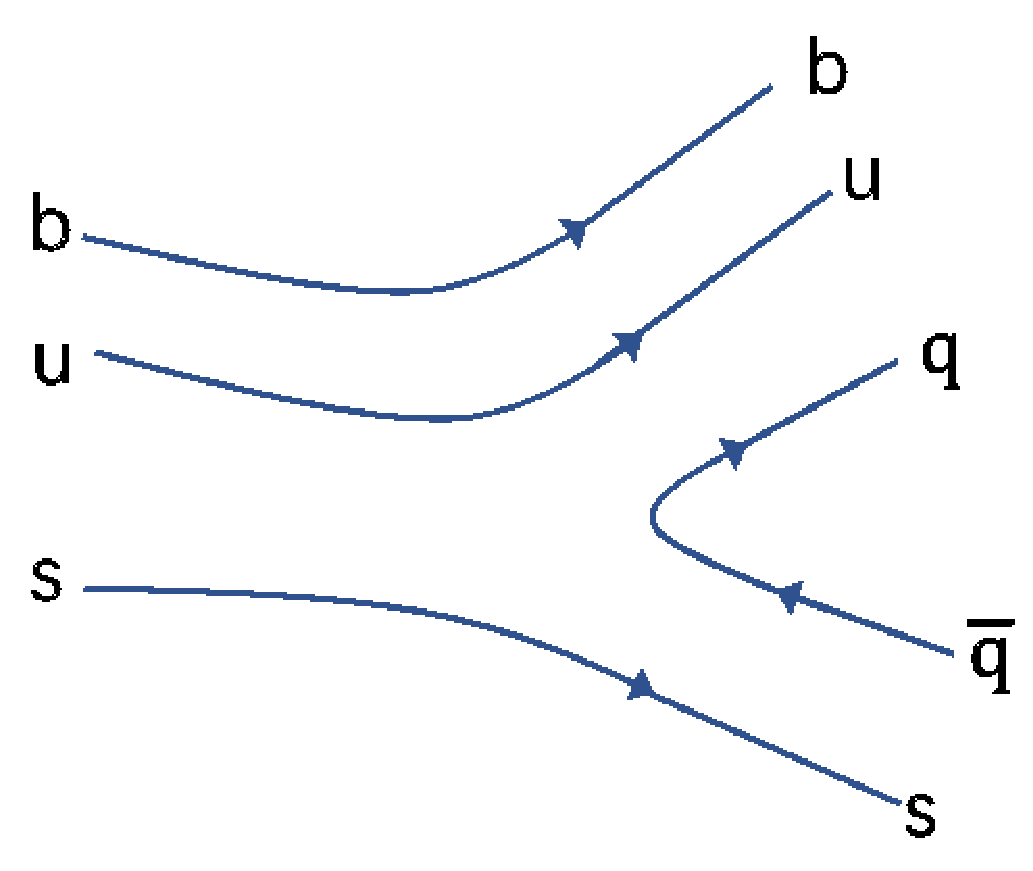}
		\end{minipage}}
		\centering
		\caption{Decay process of $\Xi^{(')}_b$ in QPC model.}\label{FIG1}
	\end{center}
\end{figure}
Here, we adopt this model to study the strong decays of $\Xi^{(')}_b$ baryons.
The  $\Xi^{(')}_b$ is made of two light flavor quarks(u,s) and a heavy flavor quark b.
According to the quark rearrangement process, any of the
three quarks in the initial baryon can go into the final meson. Hence, as shown in Fig~\ref{FIG1}, there are three possible decay ways. Meanwhile, the possible final states of  $\Xi_b$ and $\Xi_b^{'}$ are listed in table~\ref{table1}.

In the nonrelativistic limit, the transition operator under the
QPC model is given by

\begin{equation}
	\begin{aligned}
		T=&-3\gamma\sum\limits_{m}\langle1m1-m|00\rangle\int d^3\textbf{p}_4d^3\textbf{p}_5\delta^3(\textbf{p}_4+\textbf{p}_5)\\
		&\times\mathcal{Y}^m_1(\dfrac{\textbf{p}_4-\textbf{p}_5}{2})\chi^{45}_{1-m}\phi^{45}_0\omega^{45}_0a^\mathcal{y}_{4i}(\textbf{p}_4)b^\mathcal{y}_{5j}(\textbf{p}_5),\\
	\end{aligned}
\end{equation}
where $\gamma$ is a parameter of vacuum pair-production strength.
$\textbf{p}_4$ and $\textbf{p}_5$ present the momenta of the two created quarks, respectively.
$\omega^{45}_0$=$\delta_{ij}$ ,
$\phi^{45}_0$=(u$\rm{\overline{u}}$+d$\rm{\overline{d}}$+s$\rm{\overline{s}}$)/$\sqrt{3}$ and $\chi^{45}_{1-m}$ mean the wave function of quark pair flavor singlet, color singlet and spin triplet state.
The solid harmonic polynomial $\mathcal{Y}^m_1$=$|\textbf{p}|$$\rm{\boldsymbol{Y}}^m_1$($\theta_p\phi_p$) reflects the momentum-space distribution and
the creation operatora $a^\mathcal{y}_{4i}d^\mathcal{y}_{5j}$ denotes the quark pair-creation in the vacuum.

According to the definition of the mock state in the constituent quark model~\cite{Hayne:1981zy}, the spatial wave functions of the  initial baryon A or final baryon B is
\begin{equation}
	\begin{aligned}
		&|A(N^{2S_A+1}_AL_AJ_AM_{J_A})(\textbf{p}_A)\rangle\\
		=&\sqrt{2E_A}\varphi^{123}_A\omega^{123}_A\sum\limits_{M_{L_A},M_{S_A}}\langle L_AM_{L_A};S_AM_{S_A}|J_A,M_{J_A}\rangle\\
		&\times\int d^3\textbf{p}_1d^3\textbf{p}_2d^3\textbf{p}_3\delta^3(\textbf{p}_1+\textbf{p}_2+\textbf{p}_3-\textbf{p}_A)\\
		&\times\Psi_{N_AL_AM_{L_A}(\textbf{p}_1\textbf{p}_2\textbf{p}_3)}\chi^{123}_{S_AM_{S_A}}|q_1(\textbf{p}_1)q_2(\textbf{p}_2)q_3(\textbf{p}_3)\rangle.,
	\end{aligned}
\end{equation}

Similarly,  the spatial wave function of  final meson C is
\begin{equation}
	\begin{aligned}
		&|C(N^{2S_C+1}_CL_CJ_CM_{J_C})(\textbf{p}_C)\rangle\\
		=&\sqrt{2E_C}\varphi^{ab}_C\omega^{ab}_C\sum\limits_{M_{L_C}M_{S_C}}\langle L_CM_{L_C};S_CM_{S_C}|J_CM_{J_C}\rangle\\
		&\times\int d^3\textbf{p}_ad^3\textbf{p}_b\delta^3(\textbf{p}_a+\textbf{p}_b-\textbf{p}_C)\\
		&\times\Psi_{N_CL_CM_{L_C}(\textbf{p}_a\textbf{p}_b)}\chi^{ab}_{S_CM_{S_C}}|q_a(\textbf{p}_a)q_b(\textbf{p}_b)\rangle.	
	\end{aligned}
\end{equation}
The $\textbf{p}_i$ (i = 1, 2, 3 and a, b) denotes the momentum of quarks
in hadron A and C. $\textbf{P}_A$ and $\textbf{P}_C $ are the momenta of initial baryon A and final meson C, respectively.
In the present work, the spatial wave function of a baryon or meson is described with simple harmonic oscillator space-wave function. Hence, the spatial wave function of baryon A or B without the radial excitation is
\begin{equation}
	\begin{aligned}
	&\psi(l_{\rho},m_{\rho},l_{\lambda},m_{\lambda})\\
	=&3^\frac{3}{4}(-i)^l_{\rho}\left[\dfrac{2^{l_{\rho}+2}}{\sqrt{\pi}(2l_{\rho}+1)!!}\right]^\frac{1}{2}\left(\dfrac{1}{\alpha}\right)^{l_{\rho}+\frac{3}{2}}{\rm {exp}}(-\dfrac{\textbf{p}^2_{\rho}}{2\alpha^2_{\rho}})\mathcal{Y}^{m_{\rho}}_{l_{\rho}}(\textbf{p}_{\rho})\\
	&\times(-i)^l_{\lambda}\left[\dfrac{2^{l_{\lambda}+2}}{\sqrt{\pi}(2l_{\lambda}+1)!!}\right]^\frac{1}{2}\left(\dfrac{1}{\alpha}\right)^{l_{\lambda}+\frac{3}{2}}{\rm {exp}}(-\dfrac{\textbf{p}^2_{\lambda}}{2\alpha^2_{\lambda}})\mathcal{Y}^{m_{\lambda}}_{l_{\lambda}}(\textbf{p}_{\lambda}).
	\end{aligned}
\end{equation}
The ground state spatial wave function of  meson  C is
\begin{equation}
	\psi_{0,0}=\left(\dfrac{R^2}{\pi}\right)^\frac{3}{4}{\rm {exp}}\left(-\dfrac{R^2\textbf{p}^2_{ab}}{2}\right).
\end{equation}
$\textbf{p}_{ab}$ stands for the relative momentum between the
quark and antiquark in the meson. Then, the partial decay amplitude in the center of mass frame can be obtained
\begin{equation}
	\begin{aligned}
		&\mathcal{M}^{M_{J_A}M_{J_B}M_{J_C}}(A\rightarrow B+C)\\
		=&\gamma\sqrt{8E_AE_BE_C}\prod_{A,B,C}\langle\chi^{124}_{S_BM_{S_B}}\chi^{35}_{S_CM_{S_C}}|\chi^{123}_{S_AM_{S_A}}\chi^{45}_{1-m}\rangle\\
		&\langle\varphi^{124}_B\varphi^{35}_C|\varphi^{123}_A\varphi^{45}_0\rangle I^{M_{L_A},m}_{M_{L_B},M_{L_C}}(\textbf{p}),
	\end{aligned}
\end{equation}
where $I^{M_{L_A},m}_{M_{L_B},M_{L_C}}(\textbf{p})$ is the abbreviation of spatial integral. The Clebsch-Gorden coefficients $\prod_{A,B,C}$  can be expanded as
\begin{equation}
	\begin{aligned}
		\sum&\langle L_AM_{L_B};S_BM_{S_B}|J_B,M_{J_B}\rangle \langle L_AM_{L_C};S_CM_{S_C}|J_C,M_{J_C}\rangle\\
		&\times \langle L_AM_{L_A};S_AM_{S_A}|J_A,M_{J_A}\rangle \langle1m;1-m|00\rangle.\\
	\end{aligned}
\end{equation}
Eventually, the decay width of initial state A to final state B and C can be calculated by the following formula,
\begin{equation}
	\Gamma(A\rightarrow BC)=\pi^2 \dfrac{|\textbf{p}|}{M^2_A}\dfrac{1}{2J_A+1}\sum_{M_{J_A},M_{J_B},M_{J_C}}|\mathcal{M}^{M_{J_A},M_{J_B},M_{J_C}}|^2.
\end{equation}
$\textbf{P}$ is the momentum of the  baryon B in the center of mass frame of the  baryon A, which can be achieved by
\begin{equation}
	|\textbf{p}|=\dfrac{\sqrt{[M_A^2-(M_B-M_C)^2][M_A^2-(M_B+M_C)^2]}}{2M_A}.
\end{equation}

It should be remarked that according to our previous works, the physical properties of the singly heavy baryons tend to be better interpreted  in $j-j$ coupling scheme than L-S scheme. Thus, in this work we study the strong
decay properties of $\Xi^{(')}_b$ resonances within this coupling scheme. In
the heavy quark symmetry limit,  the states within
the $j- j$ coupling scheme are constructed by
\begin{equation}
	|J^P,j\rangle=	\left|\left\{[(l_{\rho}l_{\lambda})_L s_{\rho}]_j s_Q\right\}_{J^P}
	\right \rangle.
\end{equation}
The states within the the
$j-j$ coupling scheme can be expressed as linear combinations
of the states within the $L-S$ coupling,
\begin{equation}
	\begin{aligned}
		\left|\left\{[(l_{\rho}l_{\lambda})_L s_{\rho}]_j s_Q\right\}_{J^P}
		\right \rangle&=(-1)^{L+s_{\rho}+\frac{1}{2}+J}\sqrt{2J+1}\sum_{S}\sqrt{2S+1}\\
		&\begin{pmatrix}
			L&s_{\rho}&j\\s_Q&J&S
		\end{pmatrix}	\left|\left\{[(l_{\rho}l_{\lambda})_L (s_{\rho}s_Q)_S]_J \right\}
		\right\rangle.
	\end{aligned}
\end{equation}
In the expression, $l_{\rho}$ and $l_{\lambda}$ are the $\rho-$ and $\lambda-$modes quantum numbers of the orbital angular, respectively. The total orbital angular momentum $L$ = $|l_{\rho}-l_{\lambda}|,\cdots , l_{\rho}+l_{\lambda} $. $s_{\rho}$ is the quantum numbers of the
total spin of the two light quarks and $s_Q$ is the spin of the
heavy quark. The total spin angular momentum $S$ = $|s_{\rho}-s_Q|,\cdots , s_{\rho}+s_Q $. $J$ is the total angular momentum.
For $1D$ $\Xi_b$ states with $S$= 1/2, the relationship can be expressed as
\begin{equation}
	\Xi_b\left|J^P=\frac{3}{2}^+,2\right\rangle_{\lambda\lambda(\rho\rho)}=\left|\Xi_b~^2D_{\lambda\lambda(\rho\rho)}\ \frac{3}{2}^+\right\rangle,
\end{equation}
\begin{equation}
	\Xi_b\left|J^P=\frac{5}{2}^+,2\right\rangle_{\lambda\lambda(\rho\rho)}=\left|\Xi_b~^2D_{\lambda\lambda(\rho\rho)}\ \frac{5}{2}^+\right\rangle,
\end{equation}
For $1D$ $\Xi_b^{'}$ with $S$= 1/2 or 3/2, the relationship can be expressed as

\begin{equation}
	\Xi_b^{'}\left|J^P=\dfrac{1}{2}^+,1\right\rangle_{\lambda\lambda(\rho\rho)}=\left|\Xi_b^{'}~^4D_{\lambda\lambda(\rho\rho)}\ \dfrac{1}{2}^+\right\rangle,
\end{equation}
\begin{equation}
	\Xi_b^{'}\left|J^P=\dfrac{3}{2}^+,1\right\rangle_{\lambda\lambda(\rho\rho)}=\sqrt{\dfrac{1}{2}}\left|\Xi_b^{'}~^4D_{\lambda\lambda(\rho\rho)}\ \dfrac{3}{2}^+\right\rangle -\sqrt{\dfrac{1}{2}}\left|\Xi_b^{'}~^2D_{\lambda\lambda(\rho\rho)}\ \dfrac{3}{2}^+\right\rangle,
\end{equation}
\begin{equation}
	\Xi_b^{'}\left|J^P=\dfrac{3}{2}^+,2\right\rangle_{\lambda\lambda(\rho\rho)}=\sqrt{\dfrac{1}{2}}\left|\Xi_b^{'}~^4D_{\lambda\lambda(\rho\rho)}\ \dfrac{3}{2}^+\right\rangle +\sqrt{\dfrac{1}{2}}\left|\Xi_b^{'}~^2D_{\lambda\lambda(\rho\rho)}\ \dfrac{3}{2}^+\right\rangle,
\end{equation}
\begin{equation}
	\Xi_b^{'}\left|J^P=\dfrac{5}{2}^+,2\right\rangle_{\lambda\lambda(\rho\rho)}=\dfrac{\sqrt{7}}{3}\left|\Xi_b^{'}~^4D_{\lambda\lambda(\rho\rho)}\ \dfrac{3}{2}^+\right\rangle -\dfrac{\sqrt{2}}{3}\left|\Xi_b^{'}~^2D_{\lambda\lambda(\rho\rho)}\ \dfrac{3}{2}^+\right\rangle,
\end{equation}\\
\begin{equation}
	\Xi_b^{'}\left|J^P=\dfrac{5}{2}^+,3\right\rangle_{\lambda\lambda(\rho\rho)}=\dfrac{\sqrt{2}}{3}\left|\Xi_b^{'}~^4D_{\lambda\lambda(\rho\rho)}\ \dfrac{3}{2}^+\right\rangle +\dfrac{\sqrt{7}}{3}\left|\Xi_b^{'}~^2D_{\lambda\lambda(\rho\rho)}\ \dfrac{3}{2}^+\right\rangle,
\end{equation}
\begin{equation}
	\Xi_b^{'}\left|J^P=\dfrac{7}{2}^+,3\right\rangle_{\lambda\lambda(\rho\rho)}=\left|\Xi_b^{'}~^4D_{\lambda\lambda(\rho\rho)}\ \dfrac{7}{2}^+\right\rangle.
\end{equation}
In our calculation, we adopt $m_u$ = $m_d$ = 330 MeV , $m_s$ = 450 MeV , and $m_b$ = 5000 MeV  for the constituent quark mass. The value of the harmonic oscillator strength $R$ is 2.5 $\rm{GeV^{-1}}$ for all light flavor mesons, while it is R = 1.67 $\rm{GeV^{-1}}$ for the $B$ meson~\cite{Godfrey:2015dva}. The parameter $\alpha_\rho$ of the $\rho$-mode excitation between $u/d$ and $s$ quarks is taken as $\alpha_\rho$ = 0.42 GeV. Another harmonic oscillator parameter $\alpha_\lambda$ is obtained by~\cite{Zhong:2007gp}
\begin{equation}
	\alpha_\lambda=\left(\dfrac{3m_3}{m_1+m_2+m_3}\right)^\frac{1}{4}\alpha_\rho.
\end{equation}
The value of vacuum pair-production strength $\gamma$ is set as 6.95, consistent with our previous work. For the
strange quark pair $s\bar{s}$ creation, we use $\gamma_{s\bar{s}}$=6.95/$\sqrt{3}$.
Adittionally, the masses of final baryons and mesons derive from Particle Data Group~\cite{Workman:2022ynf}.

\section{Calculations and Results }\label{results}

In this work, we present a systematic study of the $1D$-wave $\Xi_b^{(')}$ baryons' strong decay within the $j-j$ coupling scheme
in the framework of the QPC model.
Both $\lambda$-mode and $\rho$-mode excitations are taken into consideration. Meanwhile,
by comparing the calculated results with experimental information, we also give the possible explanations of the newly observed states $\Xi_b(6327)^{0}$ and $\Xi_b(6333)^{0}$.
Moreover, we hope the theoretical predictions could provide useful information in searching for the missing $1D$-wave $\Xi_b^{(')}$ states.

%Theoretical results are presented as follows.
%In Ref~\cite{Wang:2022zcy}, the decay properties
%were also studied within the chiral quark model, for a comparison, the  results are listed in Table~\ref{table2},~\ref{table3}.
\subsection{The $1D$-wave $\Xi_b$ states}

According to the symmetry of wave functions, there are four $1D$-wave $\Xi_b$ baryons(see Table~\ref{table1}): two $\lambda$-mode states $\Xi_b|J^P=\frac{3}{2}^+,2\rangle_{\lambda\lambda}$ and $\Xi_b|J^P=\frac{5}{2}^+,2\rangle_{\lambda\lambda}$, and two $\rho$-mode states $\Xi_b|J^P=\frac{3}{2}^+,2\rangle_{\rho\rho}$ and $\Xi_b|J^P=\frac{5}{2}^+,2\rangle_{\rho\rho}$. In the following we will discuss their strong decay properties carefully.

\subsubsection{$\lambda$-mode  excitations}
\begin{figure}[h]
	\centering \epsfxsize=8.5 cm \epsfbox{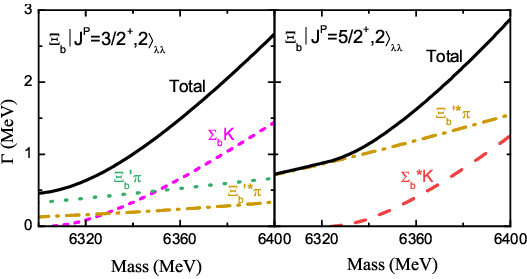}
	\caption{ Partial and total strong decay widths of the $\lambda$-mode $1D$ $\Xi_b$ states as a function of the masses. Some decay channels are not shown in the figure for their small partial decay widths.}
	\label{FIG2}
\end{figure}

As shown in Table \ref{table1}, the masses of two $1D$ $\lambda$-mode $\Xi_b$ baryons $\Xi_b|J^P=\frac{3}{2}^+,2\rangle_{\lambda\lambda}$ and $\Xi_b|J^P=\frac{5}{2}^+,2\rangle_{\lambda\lambda}$ fluctuate around $\sim6.35$ GeV.
Considering the uncertainty of the mass predictions, we plot their decay widths as a function of the mass in the range of $M\sim(6.30-6.40)$ GeV in Fig.~\ref{FIG2}. From the figure, we obtain that the two $1D$ $\lambda$-mode $\Xi_b$ baryons are likely to be narrow states.
As the mass increasing within (6.30-6.40) Gev, the partial decay widths increase stably, and the total decay widths vary within the scope of $\Gamma<3$ MeV. Similar results were also obtained in the previous works~\cite{Yao:2018jmc,Chen:2019ywy,Wang:2022zcy}.

For the $J^P=3/2^+$ state $\Xi_b|J^P=\frac{3}{2}^+,2\rangle_{\lambda\lambda}$, the main decay channels are $\Xi_b'\pi$, $\Xi_b'^*\pi$ and $\Sigma_bK$. Combining the predicted mass and decay properties, $\Xi_b|J^P=\frac{3}{2}^+,2\rangle_{\lambda\lambda}$ may be a candidate of the newly observed state $\Xi_b(6327)^0$. Hence, we fix the mass of $\Xi_b|J^P=\frac{3}{2}^+,2\rangle_{\lambda\lambda}$ at $M=6327$ MeV, and collect its decays in Table~\ref{table2}. It is found that the total decay width is about
\begin{equation}
	\Gamma_{\rm{total}}\simeq 0.74\ \rm{MeV}.
\end{equation}
This value is consistent with the observation and is about less than 3.4 times of the value in Ref.~\cite{Wang:2022zcy}. Meanwhile, $\Xi_b|J^P=\frac{3}{2}^+,2\rangle_{\lambda\lambda}$ has a significant decay rate into $\Sigma_bK$, which is predicted to be
 \begin{equation}
 \frac{\Gamma[\Xi_b|J^P=3/2^+,2\rangle_{\lambda\lambda}\to  \Sigma_b K]}{\Gamma_{\rm{total}}}\sim22 \%.
 \end{equation}
 This result is in good agreement with the predictions in Ref.~\cite{Wang:2022zcy}. In addition, this state has large decay rates into the $\Xi_b'\pi$ and $\Xi_b'^{*}\pi$ channels with branching fractions of
\begin{equation}
 \frac{\Gamma[\Xi_b|J^P=3/2^+,2\rangle_{\lambda\lambda}\to  \Xi_b' \pi]}{\Gamma_{\rm{total}}}\sim55 \%,
 \end{equation}
 \begin{equation}
  \frac{\Gamma[\Xi_b|J^P=3/2^+,2\rangle_{\lambda\lambda}\to  \Xi_b'^{*} \pi]}{\Gamma_{\rm{total}}}\sim23 \%.
 \end{equation}
Those results are good agreement with the ratios in Ref.~\cite{Wang:2022zcy} as well.

If the newly observed state $\Xi_b(6327)^0$ corresponds to $\Xi_b|J^P=\frac{3}{2}^+,2\rangle_{\lambda\lambda}$ indeed, besides the $\Lambda_bK\pi$ final state by the intermediate channel $\Sigma_bK$, the $\Xi_b\pi\pi$ final state by the intermediate channels $\Xi_b'\pi$ and $\Xi_b'^{*}\pi$ may be another interesting channel for observation of $\Xi_b(6327)^0$ in forthcoming experiments.

\begin{table*}
	\setlength{\tabcolsep}{6pt}
	\caption{The comparison of the partial decay widths of the $1D$-wave  $\Xi_b$ states from the QPC model and the chiral quark model~\cite{Wang:2022zcy}. $\Gamma_{\mathrm{Total}}$ represent the total decay width and Expt. stands for the experimental value. The unit is MeV.}
	\centering
	\begin{tabular}{c|cc|cc|cc|cc}
		\hline\hline
		\multirow{3}{*}{Decay Width}&\multicolumn{2}{c}{$\Xi_{b}\ket{J^{P}=\frac{3}{2}^{+},2}_{\lambda\lambda}$}		&\multicolumn{2}{c}{$\Xi_{b}\ket{J^{P}=\frac{5}{2}^{+},2}_{\lambda\lambda}$}&\multicolumn{2}{c}{$\Xi_{b}\ket{J^{P}=\frac{3}{2}^{+},2}_{\rho\rho}$}
& \multicolumn{2}{c}{$\Xi_{b}\ket{J^{P}=\frac{5}{2}^{+},2}_{\rho\rho}$}\\
		\cline{2-9}
		&\multicolumn{2}{c}{$\Xi_b(6327)^0$}&\multicolumn{2}{c}{$\Xi_b(6333)^0$}&\multicolumn{2}{c}{M=6420}&\multicolumn{2}{c}{M=6430}\\
		\cline{2-9}
		&QPC&ChQM~\cite{Wang:2022zcy}&QPC&ChQM~\cite{Wang:2022zcy}&QPC&ChQM~\cite{Wang:2022zcy}&QPC&ChQM~\cite{Wang:2022zcy}\\
		\hline
		$\Gamma[\Sigma_{b}K]$& 0.16&0.59&...&...&17.7&8.59&0.67&0.34 \\
		$\Gamma[\Sigma^{*}_{b}K]$&...&...&0.04&0.11&2.91&1.56&18.2&9.25 \\
		$\Gamma[\Xi^{'}_{b}\pi]$& 0.41&1.30&0.02&0.41&21.6&4.63&1.57&0.95 \\
		$\Gamma[\Xi^{'*}_{b}\pi]$& 0.17&0.67&0.96&1.64&6.02&1.97&27.0&5.87 \\
		\hline
		$\Gamma[\rm{Total}]$& 0.74&2.56&1.02&2.16&48.3&16.75&47.5&16.42\\
		\hline
		Expt.&\multicolumn{2}{c}{$0.93_{-0.60}^{+0.74}$}&\multicolumn{2}{c}{$0.25_{-0.25}^{+0.58}$}&\multicolumn{2}{c}{-}&\multicolumn{2}{c}{-} \\
		\hline\hline
	\end{tabular}
	\label{table2}
\end{table*}
%The large branching fraction of $\Xi^{'}_b\pi$ and $\Xi^{'*}_b\pi$ indicates the state $\Xi_b|J^P=3/2^+,2\rangle_{\lambda\lambda}$ has a good potential to be observed in $\Xi_b\pi\pi$  final state via the decay chains $\Xi_b|J^P=3/2^+,2\rangle_{\lambda\lambda}\to \Xi^{'}_b\pi \to \Xi_b\pi\pi$ or $\Xi_b|J^P=3/2^+,2\rangle_{\lambda\lambda}\to \Xi^{'*}_b\pi \to \Xi_b\pi\pi$.

The $J^P=5/2^+$ state $\Xi_b|J^P=5/2^+,2\rangle_{\lambda\lambda}$ mainly decays into the $\Xi_b'^{*}\pi$ and $\Sigma_b^*K$ channels. The $\Xi_b|J^P=5/2^+,2\rangle_{\lambda\lambda}$ may be an assignment of the newly observed state $\Xi_b(6333)^0$, since the $\Sigma_b^*K$ mode has a significant contribution to the decay. Fixing the mass of $\Xi_b|J^P=5/2^+,2\rangle_{\lambda\lambda}$ at $M=6333$ MeV, we collect its decay properties in Table~\ref{table2} as well. From the table, we find that the total decay width
\begin{equation}
	\Gamma_{\rm{Total}}\simeq 1.02\ \rm{MeV},\\
\end{equation}
is about less than 2 times of the result in Ref.~\cite{Wang:2022zcy} and close to the upper limit of the observed one. The branching fractions for the dominant decay channels are
\begin{equation}
	\begin{aligned}
		&\dfrac{\Gamma[\Xi_b|J^P=5/2^+,2 \rangle_{\lambda\lambda}\to \Sigma^{*}_b K]}{\Gamma_{\rm{total}}}\sim4\ \%,\\
		&\dfrac{\Gamma[\Xi_b|J^P=5/2^+,2 \rangle_{\lambda\lambda}\to\Xi^{'*}_b \pi]}{\Gamma_{\rm{total}}}\sim 94\ \%.
	\end{aligned}	
\end{equation}
According to our calculations, the $\Xi_b|J^P=5/2^+,2\rangle_{\lambda\lambda}$ is almost saturated by the $\Xi^{'*}_b \pi$ decay channel. Hence, to further decode the nature of $\Xi_b(6333)^0$, the $\Xi_b\pi\pi$ final state by the intermediate channel $\Xi^{'*}_b \pi$ is worth attention.
 This result is highly agreement with the calculation in Ref.~\cite{Wang:2022zcy}.

%taking the mass as 6333 MeV, we also get its total width and the branching ratios for principal channels  as follows,
%\begin{equation}
%	\Gamma_{\rm{total}}\simeq 1.0\ \rm{MeV},\\
%\end{equation}
%\begin{equation}
%	\begin{aligned}
%		&\dfrac{\Gamma[\Xi_b|J^P=5/2^+,2 \rangle_{\lambda\lambda}\to \Sigma^{*}_b K]}{\Gamma_{\rm{total}}}\sim3.8\ \%,\\
%		&\dfrac{\Gamma[\Xi_b|J^P=5/2^+,2 \rangle_{\lambda\lambda}\to\Xi^{'}_b \pi]}{\Gamma_{\rm{total}}}\sim1.6\ \%.\\
%		&\dfrac{\Gamma[\Xi_b|J^P=5/2^+,2 \rangle_{\lambda\lambda}\to\Xi^{'*}_b \pi]}{\Gamma_{\rm{total}}}\sim 95\ \%.
%	\end{aligned}	
%\end{equation}
%It's interesting that  $\Xi^{'*}_b\pi$   even saturates $\Xi_b|J^P=5/2^+,2\rangle_{\lambda\lambda}$ decay channels.
%So it is suggested to find other siginals in the $\Xi_b\pi\pi$ final state.
%
%
%It's apparent that decay properties of $\Xi_b|J^P=3/2^+,2\rangle_{\lambda\lambda}$ and $\Xi_b|J^P=5/2^+,2\rangle_{\lambda\lambda}$ have significant difference in $\Sigma_bK$ and $\Sigma_b^{*}K$.
%What's more, $\Xi^{'}_b\pi$ and $\Xi^{'*}_b\pi$ are ideal channels for searching and distinguishing them.
%With a comprehensive analysis of masses and decay properties, $\Xi_b|J^P=3/2^+,2\rangle_{\lambda\lambda}$ is consistent with newly observed $\Xi_b(6327)^{0}$ and $\Xi_b|J^P=5/2^+,2\rangle_{\lambda\lambda}$ correspond with  $\Xi_b(6333)^{0}$.
%The conclusion is highly coincides with our group previous work in Ref.~\cite{Wang:2022zcy}.
%

\subsubsection{$\rho$-mode  excitations}
\begin{figure}[h]
	\centering \epsfxsize=8.5 cm \epsfbox{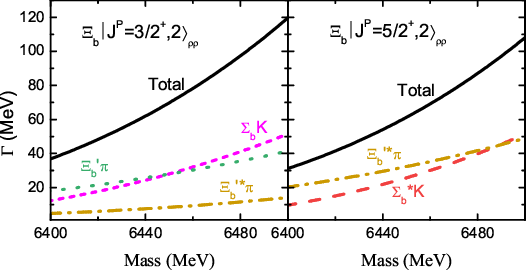}
	\caption{ Partial and total strong decay widths of the $\rho$-mode $1D$ $\Xi_b$ states as a function of the masses. Some decay channels are not shown in the figure for their small partial decay widths.}
	\label{FIG3}
\end{figure}

There are a few discussions on $\rho$-mode excitation spectrum, and just roughly get that the mass of the $\rho$-mode excitation were about 100 MeV heavier that those of $\lambda$-mode excitation~\cite{Narodetskii:2008pn}. Thus, we discuss the decay properties of the two $\rho$-mode $1D$ $\Xi_b$ states as a function of the mass within the possible range $M\sim(6.40-6.50)$ GeV in Fig.~\ref{FIG3}. As shown in the figure, the two states $\Xi_b|J^P=3/2^+,2\rangle_{\rho\rho}$ and $\Xi_b|J^P=5/2^+,2\rangle_{\rho\rho}$ are probably moderate states with a width of dozens of MeV. The $\Xi_b|J^P=3/2^+,2\rangle_{\rho\rho}$ dominantly decays into the $\Xi_b'\pi$, $\Sigma_bK$ and $\Xi_b'^*\pi$ channels. While the $\Xi_b|J^P=5/2^+,2\rangle_{\rho\rho}$ mainly decays into the $\Sigma_b^*K$ and $\Xi_b'^*\pi$ channels.
%Those decay properties are roughly consistent with the results in Ref.~\cite{Wang:2022zcy} using the chiral quark model.

To show the decay properties of the two $\rho$-mode $1D$ $\Xi_b$ states with a more comprehensible way, we further fix their masses on the predicted masses in Ref.~\cite{Yu:2021zvl} and collect their decays in Table.~\ref{table2}. Meanwhile, we also give the results from the Ref.~\cite{Wang:2022zcy} in Table.~\ref{table2} for comparison.

With the mass of $\Xi_b|J^P=3/2^+,2\rangle_{\rho\rho}$ fixed on $M=6420$ MeV, the total decay width
\begin{equation}
\Gamma_{\mathrm{Total}}\simeq48.3~\mathrm{MeV},
\end{equation}
is about 3 times of the result in Ref.~\cite{Wang:2022zcy}. The predicted partial width ratio between the mainly decay channels $\Sigma_bK$ and $\Xi_b'\pi$ is
\begin{equation}
\frac{\Gamma[\Xi_b|J^P=3/2^+,2\rangle_{\rho\rho}\rightarrow \Sigma_bK]}{\Gamma[\Xi_b|J^P=3/2^+,2\rangle_{\rho\rho}\rightarrow \Xi_b'\pi]}\simeq0.82.
\end{equation}

The other $\rho$-mode $1D$ $\Xi_b$ state $\Xi_b|J^P=5/2^+,2\rangle_{\rho\rho}$ has a comparative decay width with the $\Xi_b|J^P=3/2^+,2\rangle_{\rho\rho}$ when we fix the mass on $M=6430$ MeV. While the main decay channels of the $\Xi_b|J^P=5/2^+,2\rangle_{\rho\rho}$ are $\Sigma_b^*K$ and $\Xi_b'^*\pi$. The predicted partial width ratio is
\begin{equation}
\frac{\Gamma[\Xi_b|J^P=5/2^+,2\rangle_{\rho\rho}\rightarrow \Sigma_b^*K]}{\Gamma[\Xi_b|J^P=3/2^+,2\rangle_{\rho\rho}\rightarrow \Xi_b'^*\pi]}\simeq0.67.
\end{equation}
Difference dominant decay channels can be used to distinguish the two $\rho$-mode $1D$ $\Xi_b$ states.

\subsection{The $1D$-wave $\Xi_b^{'}$ states}
Based on the classification of the quark model, there are six $\rho$-mode $1D$-wave $\Xi_b^{'}$ states and six $\lambda$-mode $1D$-wave $\Xi_b^{'}$ states. According to the theoretical predictions by various quark models(see Table~\ref{table1}), the masses of the $\lambda$-mode $1D$ $\Xi_b'$ vary in the range of $M\sim(6.40-6.50)$ GeV. Meanwhile, we notice that the mass predictions for the $\rho$-mode $1D$-wave $\Xi_b^{'}$ states are lack, which may be $\sim100$ MeV heavier than those of the $\lambda$-mode excitations. Hence, the masses of the $\rho$-mode $1D$-wave $\Xi_b^{'}$ states may be in the scope of $M\sim(6.50-6.60)$ GeV. Because of high masses of the $1D$ $\Xi_b'$ excitations, there are many new decay channels are available compared to the $1D$ $\Xi_b$ baryons, such as the decay channels containing a $1P$-wave bottom baryon or $B$ meson. While, it should be remarked that the $\Lambda/\Sigma B$ decay channel for $\rho$-mode excitations is forbidden due to the orthogonality of spatial wave function. Thus, for the $\rho$-mode $1D$-wave $\Xi_b^{'}$ excitations, we only focus on the decay channels containing light $K$ and $\pi$ mesons.
%Up to now, signals of excited $\Xi_{b}^{'}$ are still very scarce, only two 1S states being established by experiments.
%Therefore, it is of great value  to give predicitons in this regard.
%The 1D-wave $\Xi_b^{'}$ states have higher masses than $\Xi_b$, vary in 6.3-6.5 GeV.
%Because of symmetric flavor functions and higher masses, apart from  the same decay final states with 1D-wave $\Xi_b$, $\Xi_b^{'}$ have some extra decay channels, such as ground and 1P-wave $\Lambda_b K$  and so on.
%For further deciphering their structures, we  give a quantitative analysis
%of decay properties of $1D_{\lambda\lambda}$ $\Xi_b^{'}$ states with predicted masses in Ref~\cite{Ebert:2011kk}.
%Unfortunately, due to the lack of predictions toward 1D-wave $\rho$-mode excited $\Xi_b^{'}$, we fix them 100 MeV above $\lambda$-mode excitaions for reference.
%Detailed discussions can be seen as follows.

\begin{table*}[]	
		\setlength{\tabcolsep}{0.06pt}
	\caption{The comparison of the partial decay widths of the $1D$-wave  $\Xi_b'$ states from the QPC model and the chiral quark model~\cite{Wang:2022zcy}. The masses of $\lambda$-mode excitations are taken from the predictions in Ref~\cite{Ebert:2011kk}. While, the masse of the $\rho$-mode excitations are fixed on the values which are 100 MeV heavier than the corresponding masses of the $\lambda$-mode excitations. $\Gamma_{\mathrm{Total}}$ represents the total decay width and the unit is MeV.}
	\centering
	\begin{tabular}{ccccccccccccccccc}	
		\hline\hline
		\multirow{3}{*}{Decay width}&\multicolumn{2}{c}{$\underline{~~\Xi^{'}_b\ket{J^{P}=\frac{1}{2}^{+},1}_{\lambda\lambda}~~}$} &\multicolumn{2}{c}{$\underline{~~\Xi^{'}_b\ket{J^{P}=\frac{3}{2}^{+},1}_{\lambda\lambda}~~}$}
		&\multicolumn{2}{c}{$\underline{~~\Xi^{'}_b\ket{J^{P}=\frac{3}{2}^{+},2}_{\lambda\lambda}~~}$}&\multicolumn{2}{c}{$\underline{~~\Xi^{'}_b\ket{J^{P}=\frac{5}{2}^{+},2}_{\lambda\lambda}~~}$}
		&\multicolumn{2}{c}{$\underline{~~\Xi^{'}_b\ket{J^{P}=\frac{5}{2}^{+},3}_{\lambda\lambda}~~}$}&\multicolumn{2}{c}{$\underline{~~\Xi^{'}_b\ket{J^{P}=\frac{7}{2}^{+},3}_{\lambda\lambda}~~}$}\\
		&\multicolumn{2}{c}{$M$=6447}&\multicolumn{2}{c}{$M$=6459}&\multicolumn{2}{c}{$M$=6431}&\multicolumn{2}{c}{$M$=6432}&\multicolumn{2}{c}{$M$=6420}&\multicolumn{2}{c}{$M$=6414}\\
		\hline
		&QPC&ChQM~\cite{Wang:2022zcy}&QPC&ChQM~\cite{Wang:2022zcy}&QPC&ChQM~\cite{Wang:2022zcy}&QPC&ChQM~\cite{Wang:2022zcy}&QPC&ChQM~\cite{Wang:2022zcy}&QPC&ChQM~\cite{Wang:2022zcy}\\
		\hline
		$\Gamma[\Lambda_bK]$& 2.8  & 2.4   & 2.8  & 2.1  &  ...  & ...  &  1.3 & ... & 2.0 & 5.8 & 0.5 & 5.4  \\
		$\Gamma[\Sigma^{*}_bK]$& 0.7  & 1.8  & 1.9  & 4.9  & 0.5  & 1.8  & 3.1  &6.8  & ...  & 0.8  & ...  &0.1\\
		$\Gamma[\Sigma_{b}K]$& 1.6 & 4.1 & 0.4 &1.1  & 3.1 &8.4  & 1.0 & 0.5 & 3.6 & 0.4 & ... & 0.2 \\
		$\Gamma[\Xi_b\eta]$& 0.4  & 0.4 & 0.5 & 0.1 & ... & 0.1 & 0.1  & 0.9 & 0.2 & 0.2 & ... & ... \\
		$\Gamma[\Xi_b\pi]$& 4.6 & 2.2 & 4.6 & 2.0 & ... & ... &2.0  & ... & 3.2 & 9.6 & 0.5 & 9.1 \\
		$\Gamma[\Xi^{'}_b\pi]$& 1.1  &1.3  & 0.3 & 0.3 & 2.3 & 2.9 & 0.8 & 1.3 & 3.2 & 1.2 & ... & 0.6 \\
		$\Gamma[\Xi^{'*}_b\pi]$& 0.5  &0.6  & 1.3 & 1.6 & 0.5 & 2.3 & 2.6 & 4.1 & 0.1 & 2.1 & 0.1 & 0.6 \\
		$\Gamma[\Lambda B]$& 10.7  & - & 0.9 & - & 3.8 & - & ... & - & ... & - & ... & - \\
		$\Gamma[\Lambda_b\ket{J^P= 1/2^-,1}_{\lambda}K]$& 34.7  &32.3  & ... & 0.2 &...  & 1.1 & ... & 2.6 & ... & 0.1 & ... & ... \\
		$\Gamma[\Lambda_b\ket{J^P= 3/2^-,1}_{\lambda}K]$& ...  & 0.1 & 36.0 & 15.8 & ... & 1.4 & ... & ... & ... & 0.2 & - & ... \\
		$\Gamma[\Xi_b\ket{J^P= 1/2^-,1}_{\lambda}\pi]$& 52.0  &13.8  & ... & 1.1 & 0.1  & 0.4 & ... & 0.6 &0.3  & 0.5 & ... & ... \\
		$\Gamma[\Xi_b\ket{J^P= 3/2^-,1}_{\lambda}\pi]$& ...  & 0.4 & 52.3 & 10.2 & 0.1 & 1.0 & 0.1 & 0.2 & 0.1 & 0.1 & 0.2  & 1.5 \\
		$\Gamma[\Xi^{'}_b\ket{J^P= 1/2^-,0}_{\lambda}\pi]$&...  & 1.0 & ... & 0.1 & 11.2 & 0.2 & ... & 0.1 & ... & ... & ... & ... \\
		$\Gamma[\Xi^{'}_b\ket{J^P= 1/2^-,1}_{\lambda}\pi]$& 21.4  & 9.0 & ... & ... & ... & 0.1 & ... & 0.1 & 0.1 & 0.1 & 0.2 & ... \\
		$\Gamma[\Xi^{'}_b\ket{J^P= 3/2^-,1}_{\lambda}\pi]$& ...  & 0.9 & 3.0 &0.3  & 1.3 & 0.6 & 0.7 & ... & 0.7  & ... & 0.8 & ... \\
		$\Gamma[\Xi^{'}_b\ket{J^P= 3/2^-,2}_{\lambda}\pi]$& ...  & 0.3 & 1.9 &...  & 18.0 & 6.2 & 0.2 & ... & 0.2 &...  & 0.2 & ... \\
		$\Gamma[\Xi^{'}_b\ket{J^P= 5/2^-,2}_{\lambda}\pi]$& 6.1  &0.6  &1.6  &0.2  & 1.8 & 0.1 & 3.3 & 1.3 & 2.4 & ... & 0.5 & ... \\
		$\Gamma_{\text{Total}}$& 136.7  & 71.2 & 107.6 & 40.0 & 42.6 & 26.6 & 15.3 & 18.5 & 16.0 & 21.0 & 3.0 & 17.5 \\
		\hline \hline
		\multirow{3}{*}{Decay width}&\multicolumn{2}{c}{$\underline{~~\Xi^{'}_b\ket{J^{P}=\frac{1}{2}^{+},1}_{\rho\rho}~~}$} &\multicolumn{2}{c}{$\underline{~~\Xi^{'}_b\ket{J^{P}=\frac{3}{2}^{+},1}_{\rho\rho}~~}$}
		&\multicolumn{2}{c}{$\underline{~~\Xi^{'}_b\ket{J^{P}=\frac{3}{2}^{+},2}_{\rho\rho}~~}$}&\multicolumn{2}{c}{$\underline{~~\Xi^{'}_b\ket{J^{P}=\frac{5}{2}^{+},2}_{\rho\rho}~~}$}
		&\multicolumn{2}{c}{$\underline{~~\Xi^{'}_b\ket{J^{P}=\frac{5}{2}^{+},3}_{\rho\rho}~~}$}&\multicolumn{2}{c}{$\underline{~~\Xi^{'}_b\ket{J^{P}=\frac{7}{2}^{+},3}_{\rho\rho}~~}$}\\
		&\multicolumn{2}{c}{$M$=6547}&\multicolumn{2}{c}{$M$=6559}&\multicolumn{2}{c}{$M$=6531}&\multicolumn{2}{c}{$M$=6532}&\multicolumn{2}{c}{$M$=6520}&\multicolumn{2}{c}{$M$=6514}\\
		\hline &QPC&ChQM~\cite{Wang:2022zcy}&QPC&ChQM~\cite{Wang:2022zcy}&QPC&ChQM~\cite{Wang:2022zcy}&QPC&ChQM~\cite{Wang:2022zcy}&QPC&ChQM~\cite{Wang:2022zcy}&QPC&ChQM~\cite{Wang:2022zcy}\\
		\hline
		$\Gamma[\Lambda_bK]$& 192.1  & 7.6 & 203.7 & 7.2 & ... & ... & ... & ... & 22.9 & 6.7 & 21.7 & 6.5 \\
		$\Gamma[\Sigma^{*}_bK]$& 1.1  & 4.4 & 20.9 & 13.8 & 22.8 & 7.2 & 108.1 & 21.1 & 3.5 & 5.3 & 4.2 & 2.0 \\
		$\Gamma[\Sigma_bK]$& 54.5  & 11.1 & 15.1 & 2.8 & 105.9 & 24.2 & 4.3 & 1.8 & 4.0 & 1.7 & 2.0 & 0.9 \\
		$\Gamma[\Xi_b\pi]$& 259.9  & 11.0 & 275.9 & 10.6 & ... & ... & ... & ... & 26.8 & 10.5 & 25.4 & 10.1 \\
		$\Gamma[\Xi^{'}_b\pi]$& 38.4  & 4.3 & 5.2 & 1.1 & 77.5 & 9.6 & 4.0 & 2.0 & 3.9 & 2.0 & 2.1 & 1.1 \\
		$\Gamma[\Xi^{'*}_b\pi]$& 0.3  & 2.1 & 45.9 & 5.4 & 19.3 & 4.8 & 195.6 & 11.4 & 11.0 & 5.3 & 9.3 & 2.7 \\
		
		$\Gamma[\Xi_b\eta]$& 14.4  & 0.8 & 16.2 & 0.6 & ... & 1.0 & ... & 2.6 & 0.4 & 0.6 & 0.4 & ... \\
		
		$\Gamma[\Xi^{'}_b\eta]$& 0.5  & 0.6 & 0.2 & 0.7 & 0.7 & 0.2 & ... & 0.2 & ... & ... & ... & ... \\
		
		$\Gamma[\Xi^{'*}_b\eta]$& 0.1  & 0.1 & 0.6 & 0.8 & 0.1 & 0.4 & 2.3 & 0.2 & 0.1 & 0.1 & ... & ... \\
		$\Gamma[\Lambda_b\ket{J^P= 1/2^-,1}_{\lambda}K]$ &  6.4 & 0.5 & 0.6 & 0.6 & ... & ... & ... & ... & 0.4 & ... & ... & ... \\
		$\Gamma[\Lambda_b\ket{J^P= 3/2^-,1}_{\lambda}K]$ & 0.8  &0.9  & 7.2 & 1.9 & ... & 0.1 & ... & ... & 0.1 & ... & 0.5 & ... \\
		$\Gamma[\Xi_b\ket{J^P= 1/2^-,1}_{\lambda}\pi]$ & 9.8  & 0.5 & 0.9 & 0.5 & ... & ... & ... & ... & 0.8 & 0.1 & ... & 0.1 \\
		$\Gamma[\Xi_b\ket{J^P= 3/2^-,1}_{\lambda}\pi]$ & 1.3  & 0.8 & 10.8 & 1.9 & ... & 0.1 &...  & ... & 0.2 & 0.1 & 1.0 & 0.1 \\
		$\Gamma[\Xi^{'}_b\ket{J^P= 1/2^-,0}_{\lambda}\pi]$ & 0.1  & ... & 0.3 & ... & 1.6 & ... & ... & ... & ... & ... & ... & ... \\
		$\Gamma[\Xi^{'}_b\ket{J^P= 1/2^-,1}_{\lambda}\pi]$ & 4.4  & ... & 0.1 & 0.1 & ... & ... & ... & ... & ... & ... & ... & ... \\
		$\Gamma[\Xi^{'}_b\ket{J^P= 3/2^-,1}_{\lambda}\pi]$ & 0.6  & ... & 1.0 & ... & 0.5 & ... & ... & ... & ... & ... & ... & ... \\
		$\Gamma[\Xi^{'}_b\ket{J^P= 3/2^-,2}_{\lambda}\pi]$ & ...  & 0.1 & ... & 0.2 & 2.4 & ... &...  & ... & ... & ... & ... & ... \\
		$\Gamma[\Xi^{'}_b\ket{J^P= 5/2^-,2}_{\lambda}\pi]$ &  0.8 & ... & 0.4 & ... & 0.4 & ... & 1.1 & 0.1 & ... & ... & 0.1 & ... \\
		$\Gamma_{\text{Total}}$& 585.6  & 44.8 & 604.7 & 48.2 & 231.2 & 47.6 & 312.8 & 39.4 & 74.3 & 32.4 & 66.8 & 23.5 \\
		\hline \hline
	\end{tabular}	
\label{table3}
\end{table*}

\subsubsection{$\lambda$-mode  excitations}
The six $1D$-wave $\lambda$-mode excited $\Xi_b'$ states within j-j coupling scheme are $\Xi_b'|J^P=1/2^+,1\rangle_{\lambda\lambda}$, $\Xi_b'|J^P=3/2^+,1\rangle_{\lambda\lambda}$, $\Xi_b'|J^P=3/2^+,2\rangle_{\lambda\lambda}$, $\Xi_b'|J^P=5/2^+,2\rangle_{\lambda\lambda}$,
$\Xi_b'|J^P=5/2^+,3\rangle_{\lambda\lambda}$ and $\Xi_b'|J^P=7/2^+,3\rangle_{\lambda\lambda}$, respectively. Firstly, we fix their masses on the predictions within the QCD-motivated relative quark model in Ref.~\cite{Ebert:2011kk}, and collect the decay properties in  Table~\ref{table3}.

The $J^P=1/2^+$ state $\Xi_b'|J^P=1/2^+,1\rangle_{\lambda\lambda}$ may be a moderate state with a width of $\Gamma_{\mathrm{Total}}\simeq 136.7$ MeV, which is about 2 times of the value in Ref.~\cite{Wang:2022zcy}.Furthermore, we find that this state mainly decays into $\Xi_b|J^P=1/2^-,1\rangle_{\lambda} \pi$, $\Lambda_b|J^P=1/2^-,1\rangle_{\lambda} K$ and $\Xi_b'|J^P=1/2^-,1\rangle_{\lambda} \pi$. The predicted partial width ratios are
\begin{equation}
\frac{\Gamma[\Xi_b'|J^P=1/2^+,1\rangle_{\lambda\lambda}\rightarrow \Lambda_b|J^P=1/2^-,1\rangle_{\lambda}K]}
{\Gamma[\Xi_b'|J^P=1/2^+,1\rangle_{\lambda\lambda}\rightarrow \Xi_b|J^P=1/2^-,1\rangle_{\lambda} \pi]}\simeq0.67,
\end{equation}
\begin{equation}
\frac{\Gamma[\Xi_b'|J^P=1/2^+,1\rangle_{\lambda\lambda}\rightarrow \Xi_b'|J^P=1/2^-,1\rangle_{\lambda}\pi]}
{\Gamma[\Xi_b'|J^P=1/2^+,1\rangle_{\lambda\lambda}\rightarrow \Xi_b|J^P=1/2^-,1\rangle_{\lambda} \pi]}\simeq0.41.
\end{equation}
Hence, the $\Xi_b'|J^P=1/2^+,1\rangle_{\lambda\lambda}$ state are likely to be observed in the $\Xi_b\pi\pi\pi$ and $\Sigma_b\pi K$ final state via the decay chains $\Xi_b'|J^P=1/2^+,1\rangle_{\lambda\lambda}\rightarrow \Xi_b(\Xi_b')|J^P=1/2^-,1\rangle_{\lambda} \pi\rightarrow \Xi_b'\pi\pi\rightarrow \Xi_b\pi\pi\pi$ and $\Xi_b'|J^P=1/2^+,1\rangle_{\lambda\lambda}\rightarrow \Lambda_b|J^P=1/2^-,1\rangle_{\lambda} K\rightarrow \Lambda_b\pi\pi K$.

Meanwhile, the partial decay width of $\Lambda B$ is sizable, and the predicted partial width ratio between $\Lambda B$ and $\Xi_b|J^P=1/2^-,1\rangle_{\lambda} \pi$ is
\begin{equation}
\frac{\Gamma[\Xi_b'|J^P=1/2^+,1\rangle_{\lambda\lambda}\rightarrow \Lambda B]}
{\Gamma[\Xi_b'|J^P=1/2^+,1\rangle_{\lambda\lambda}\rightarrow \Xi_b|J^P=1/2^-,1\rangle_{\lambda} \pi]}\simeq0.21.
\end{equation}
$\Lambda B$ may be another good channel for looking for the missing state $\Xi_b'|J^P=1/2^+,1\rangle_{\lambda\lambda}$.

Fort the $J^P=3/2^+$ state $\Xi_b'|J^P=3/2^+,1\rangle_{\lambda\lambda}$, the width is predicted to be $\Gamma\simeq107.6$ MeV. The dominant decay channels are $\Xi_b|J^P=3/2^-,1\rangle_{\lambda} \pi$ and $\Lambda_b|J^P=3/2^-,1\rangle_{\lambda} K$, and the partial width ratio is
\begin{equation}
\frac{\Gamma[\Xi_b'|J^P=3/2^+,1\rangle_{\lambda\lambda}\rightarrow \Lambda_b|J^P=3/2^-,1\rangle_{\lambda}K]}
{\Gamma[\Xi_b'|J^P=3/2^+,1\rangle_{\lambda\lambda}\rightarrow \Xi_b|J^P=3/2^-,1\rangle_{\lambda} \pi]}\simeq0.69.
\end{equation}
This state has good potential to be discovered in the $\Xi_b\pi\pi\pi$ and $\Lambda_b\pi\pi K$ final states via the decay chains $\Xi_b^{'}|J^P=3/2^+,1\rangle_{\lambda\lambda}\to\Xi_b\ket{J^P= 3/2^-,1}_{\lambda}\pi\to\Xi^{'*}_{b}\pi
\pi\to\Xi_b\pi\pi\pi$ and $\Xi_b^{'}|J^P=3/2^+,1\rangle_{\lambda\lambda}\to\Lambda_b\ket{J^P= 3/2^-,1}_{\lambda}K\rightarrow \Lambda_b\pi\pi K$. The predicted dominant decay channels are consistent with the results within the ChQM.

The other $J^P=3/2^+$ state $\Xi_b'|J^P=3/2^+,2\rangle_{\lambda\lambda}$ has a width of $\Gamma\simeq42.6$ MeV, and mainly decays into $\Xi_b'\ket{J^P= 3/2^-,2}_{\lambda}\pi$ and $\Xi_b'\ket{J^P= 1/2^-,0}_{\lambda}\pi$ with a partial width ratio
\begin{equation}
\frac{\Gamma[\Xi_b'|J^P=3/2^+,2\rangle_{\lambda\lambda}\rightarrow \Xi_b'\ket{J^P= 1/2^-,0}_{\lambda}\pi]}
{\Gamma[\Xi_b'|J^P=3/2^+,1\rangle_{\lambda\lambda}\rightarrow \Xi_b'\ket{J^P= 3/2^-,2}_{\lambda}\pi]}\simeq0.62.
\end{equation}
Since the predicted width of $\Xi_b'|J^P=3/2^+,2\rangle_{\lambda\lambda}$ is not broad, this resonance might be observed via the decays chains
$\Xi_b^{'}|J^P=3/2^+,2\rangle_{\lambda\lambda}\to\Xi^{'}_b\ket{J^P= 3/2^-,2}_{\lambda}\pi\to\Xi_b\pi\pi$ and $\Xi_b^{'}|J^P=3/2^+,2\rangle_{\lambda\lambda}\to\Xi^{'}_b\ket{J^P= 1/2^-,0}_{\lambda}\pi\to\Xi_b \pi\pi$.

Furthermore, $\Xi_b'|J^P=3/2^+,2\rangle_{\lambda\lambda}$ may have a sizeable decay rate into $\Lambda B$, and predicted branching fractions is
\begin{equation}
\frac{\Gamma[\Xi_b'|J^P=3/2^+,2\rangle_{\lambda\lambda}\rightarrow \Lambda B]}
{\Gamma_{\mathrm{Total}}}\sim9\%.
\end{equation}
Thus, $\Lambda B$ may be another interesting channel for experimental exploration.

The two $J^P=5/2^+$ states $\Xi_b'|J^P=5/2^+,2\rangle_{\lambda\lambda}$ and $\Xi_b'|J^P=5/2^+,3\rangle_{\lambda\lambda}$ have a comparable narrow width of $\Gamma\simeq(15-16)$ MeV. Meanwhile, both of the two states have significant partial widths of the $\Xi_b'\ket{J^P= 5/2^-,2}_{\lambda}\pi$ and $\Xi_b\pi$ channels, and the corresponding branching fractions are
\begin{equation}
\frac{\Gamma[\Xi_b'|J^P=5/2^+,2/3\rangle_{\lambda\lambda}\rightarrow \Xi'_b\ket{J^P= 5/2^-,2}_{\lambda}\pi]}
{\Gamma_{\mathrm{Total}}}\sim22/15\%,
\end{equation}
\begin{equation}
\frac{\Gamma[\Xi_b'|J^P=5/2^+,2/3\rangle_{\lambda\lambda}\rightarrow \Xi_b\pi]}
{\Gamma_{\mathrm{Total}}}\sim13/20\%.
\end{equation}
In addition, the $\Xi_b'|J^P=5/2^+,2\rangle_{\lambda\lambda}$ has sizeable decay rates into $\Sigma_b^*K$ and $\Xi_b'^*\pi$, which are
\begin{equation}
\frac{\Gamma[\Xi_b'|J^P=5/2^+,2\rangle_{\lambda\lambda}\rightarrow \Sigma_b^*K/\Xi_b'^*\pi]}
{\Gamma_{\mathrm{Total}}}\sim20/17\%.
\end{equation}
While, the $\Xi_b'|J^P=5/2^+,3\rangle_{\lambda\lambda}$ has sizeable decay rates into $\Sigma_bK$ and $\Xi_b'\pi$, and the predicted branching fractions are
\begin{equation}
\frac{\Gamma[\Xi_b'|J^P=5/2^+,3\rangle_{\lambda\lambda}\rightarrow \Sigma_bK/\Xi_b'\pi]}
{\Gamma_{\mathrm{Total}}}\sim23/20\%.
\end{equation}
Thanks to the huge difference in branching ratios, those strong decay channels can be used to distinguish the two $J^P=5/2^+$ states.

\begin{figure*}
	\centering \epsfxsize=15.0 cm \epsfbox{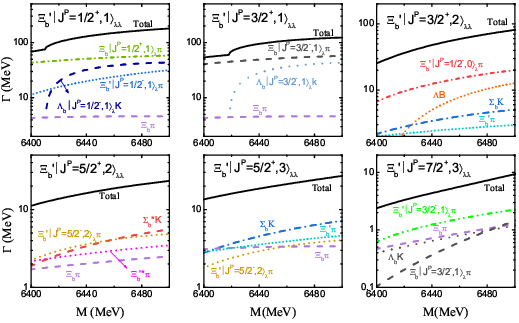}
	\caption{ Partial and total strong decay widths of the $\lambda$-mode $1D$ $\Xi_b'$ states as a function of the masses. Some decay channels are not shown in the figure for their small partial decay widths.}
	\label{FIG4}
\end{figure*}
\begin{figure*}
	\centering \epsfxsize=15.0 cm \epsfbox{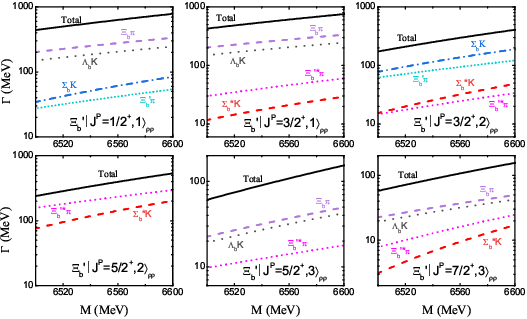}
	\caption{ Partial and total strong decay widths of the $\rho$-mode $1D$ $\Xi_b'$ states as a function of the masses. Some decay channels are not shown in the figure for their small partial decay widths.}
	\label{FIG5}
\end{figure*}

For the $J^P$=$7/2^+$ state $\Xi^{'}_b|J^P=7/2^+,3\rangle_{\lambda\lambda}$, it may be a fairly narrow state with a width of $\Gamma\simeq3$ MeV, which is significantly smaller than the results within the ChQM. Furthermore, the strong decay is mainly contributed by $\Lambda_bK$, $\Xi^{'}_{b}\pi$, $\Xi^{'}_{b}\ket{J^P= 3/2^-,1}\pi$ and $\Xi^{'}_{b}\ket{J^P= 5/2^-,2}\pi$, and each channel takes over about $17\sim27\ \%$.
So this state has a good chance to be observed in the $\Lambda_bK$, $\Xi_b\pi\pi$ and $\Xi_b\pi\pi\pi$ final states in future experiments.

Then accounting for the uncertainty of the predicted masses, we plot the decay properties of the $1D$ $\lambda$-mode $\Xi^{'}_{b}$ as functions of masses within the range of $M=(6.40-6.50)$ GeV, as shown in Fig~\ref{FIG4}.
We can easily find that partial decay widths of these states are sensitive to masses, especially for the final states containing a $P$-wave baryon.

\subsubsection{$\rho$-mode  excitations}

The discussions about the decay and spectrum of the $\rho$-mode excitations are scarce.
Considering the masses of the $\rho$-mode excitations being about
70-150 MeV higher than the $\lambda$-mode excitations, hence,
we fix the masses of the $1D$-wave $\rho$-mode excitations on the values which are 100 MeV higher than the predicted masses of the corresponding $\lambda$-mode excitations in Ref~\cite{Ebert:2011kk}. Then we
  comprehensively analyze their decay properties and collect in Table.~\ref{table3} as well.
%It should be mentioned that the spatial wave function of hadrons is adopted the harmonic oscillator form in this work, the  decay channels containing B mesons for the $\rho$-mode excited $\Xi^{'}_{b}$ are forbidden. Hence, for the $\rho$-mode excitations, we only focus on
%the decay channels containing light flavor mesons.

From the table, it is found that the two states $\Xi^{'}_b|J^P=1/2^+,1\rangle_{\rho\rho}$ and $\Xi^{'}_b|J^P=3/2^+,1\rangle_{\rho\rho}$ may be very broad states with a comparable width of $\Gamma\sim600$ MeV, and both mainly decay into the $\Lambda_bK$ and $\Xi_b\pi$ channels. The predicted branching fractions are
\begin{equation}
\dfrac{\Gamma[\Xi_b^{'}|J^P=1/2^+,1 \rangle_{\rho\rho}\to\Xi_b\pi/\Lambda_bK]}{\Gamma_{\rm{total}}}\sim44/33\%,
\end{equation}
\begin{equation}
	\dfrac{\Gamma[\Xi_b^{'}|J^P=3/2^+,1 \rangle_{\rho\rho}\to\Xi_b\pi/\Lambda_bK]}{\Gamma_{\rm{total}}}\sim46/34\%.
\end{equation}
We notice that the total decay widths of the two states in QPC are much larger than the predictions($\Gamma\sim50$ MeV) using the ChQM, while the main decay channels are grossly consistent within the two models.

Meanwhile, the partial decay widths of $\Xi^{'}_b|J^P=1/2^+,1\rangle_{\rho\rho}$ decaying into $\Sigma_bK$ and $\Xi^{'}_b|J^P=1/2^+,1\rangle_{\rho\rho}$ decaying into $\Xi_b'^*\pi$ are sizable, and the branching fractions are
\begin{equation}
\dfrac{\Gamma[\Xi_b^{'}|J^P=1/2^+,1 \rangle_{\rho\rho}\to\Sigma_bK]}{\Gamma_{\rm{total}}}\sim9\%,
\end{equation}
\begin{equation}
	\dfrac{\Gamma[\Xi_b^{'}|J^P=3/2^+,1 \rangle_{\rho\rho}\to\Xi_b'^*\pi]}{\Gamma_{\rm{total}}}\sim8\%.
\end{equation}
Anyway, it is hard to observe the two states $\Xi^{'}_b|J^P=1/2^+,1\rangle_{\rho\rho}$ and $\Xi^{'}_b|J^P=3/2^+,1\rangle_{\rho\rho}$ for their broad decay widths. Meanwhile, since they are very similar decay properties, it will be a great challenge to distinguish them from each other.

The two states $\Xi^{'}_b|J^P=3/2^+,2\rangle_{\rho\rho}$ and $\Xi^{'}_b|J^P=5/2^+,2\rangle_{\rho\rho}$ may be broad states as well, and have a comparable width of about $\Gamma\sim(230-320)$ MeV. However, our calculations indicate that their dominant decay channels are drastically different. The state $\Xi^{'}_b|J^P=3/2^+,2\rangle_{\rho\rho}$ mainly decays into $\Sigma_bK$ and $\Xi_b'\pi$, while $\Xi^{'}_b|J^P=5/2^+,2\rangle_{\rho\rho}$ decays primarily through the $\Sigma_b^*K$ and $\Xi_b'^*\pi$ channels. The predicted branching fractions are
\begin{equation}
\dfrac{\Gamma[\Xi_b^{'}|J^P=3/2^+,2 \rangle_{\rho\rho}\to\Sigma_bK/\Xi_b'\pi]}{\Gamma_{\rm{total}}}\sim46/33\%,
\end{equation}
\begin{equation}
	\dfrac{\Gamma[\Xi_b^{'}|J^P=5/2^+,2 \rangle_{\rho\rho}\to\Sigma_b^*K/\Xi_b'^*\pi]}{\Gamma_{\rm{total}}}\sim35/63\%.
\end{equation}
Comparing to the results within ChQM, we also get that the total decay widths of the two states $\Xi^{'}_b|J^P=3/2^+,2\rangle_{\rho\rho}$ and $\Xi^{'}_b|J^P=5/2^+,2\rangle_{\rho\rho}$ using QPC are significant larger, while the main decay channels predicted by the two models are consistent.

  As to the states $\Xi^{'}_b|J^P=5/2^+,3\rangle_{\rho\rho}$ and $\Xi^{'}_b|J^P=7/2^+,3\rangle_{\rho\rho}$, they have extremely similar decay properties. Both of them are probably two moderate states with a width of $\Gamma\sim$ (65-75) MeV, and predominantly decays into $\Xi_{b}\pi$, $\Lambda_bK$ and $\Xi^{'*}_{b}\pi$. The branching fractions are
 \begin{equation}
\dfrac{\Gamma[\Xi_b^{'}|J^P=5/2^+,3 \rangle_{\rho\rho}\to\Xi_{b}\pi/\Lambda_bK/\Xi^{'*}_{b}\pi]}{\Gamma_{\rm{total}}}\sim36/31/15\%,
\end{equation}
\begin{equation}
	\dfrac{\Gamma[\Xi_b^{'}|J^P=7/2^+,3 \rangle_{\rho\rho}\to\Xi_{b}\pi/\Lambda_bK/\Xi^{'*}_{b}\pi]}{\Gamma_{\rm{total}}}\sim38/32/14\%.
\end{equation}
The results can be tested in future experiments. Meanwhile, Thanks to the width not very wide, $\Xi^{'}_b|J^P=5/2^+,3\rangle_{\rho\rho}$ and $\Xi^{'}_b|J^P=7/2^+,3\rangle_{\rho\rho}$ are suggested to be searched in the $\Xi_{b}\pi,\Lambda_bK$ and $\Xi^{'*}_{b}\pi$ decay channels.

  %seem promising to be discovered.
% However, both of them predominantly decays into $\Xi_{b}\pi,\Lambda_bK$ and $\Xi^{'*}_{b}\pi$, which brought us difficulty identifying them.

Similarly, the predicted masses of the $\rho$-mode $\Xi_b'$ excitations certainly have a large uncertainty, which may bring uncertainties to our theoretical results. To investigate this effect, we plot the partial decay widths of the $\rho$-mode excitations as  functions of the masses in the range of $M=(6.50-6.60)$ GeV, as shown in Fig.~\ref{FIG5}.

\section{summary}\label{summary}
To decode the inner structures of the newly observed single bottom states, $\Xi_b(6327)^0$ and $\Xi_b(6333)^0$, we perform a systematical study of the $1D$-wave $\Xi_b$ and $\Xi_b'$ baryons in the framework of the QPC model within the $j-j$ coupling scheme. Both the $\lambda$-mode and $\rho$-mode excitations are taken into account. Our main results are summarized as follows.

For the two $1D$ $\lambda$-mode $\Xi_b$ baryons, $\Xi_b|J^P=\frac{3}{2}^+,2\rangle_{\lambda\lambda}$ and $\Xi_b|J^P=\frac{5}{2}^+,2\rangle_{\lambda\lambda}$, they should be very narrow states with a width of $\Gamma\sim1.0$ MeV. Combining with the observations, $\Xi_b|J^P=\frac{3}{2}^+,2\rangle_{\lambda\lambda}$ may be a good candidate of the $\Xi_b(6327)^0$, and $\Xi_b|J^P=\frac{5}{2}^+,2\rangle_{\lambda\lambda}$ could be a good candidate of the $\Xi_b(6333)^0$. If the assignment is true indeed, another interesting channel for observation of $\Xi_b(6327)^0$ in future experiments is $\Xi_b'\pi$, and that of $\Xi_b(6333)^0$ is $\Xi_b'^*\pi$.

The $1D$ $\rho$-mode $\Xi_b$ baryons, $\Xi_b|J^P=\frac{3}{2}^+,2\rangle_{\rho\rho}$ and $\Xi_b|J^P=\frac{5}{2}^+,2\rangle_{\rho\rho}$, have a comparable decay width of about $\Gamma\sim50$ MeV. Menawhile, $\Xi_b|J^P=\frac{3}{2}^+,2\rangle_{\rho\rho}$ mainly decays into $\Sigma_bK$ and $\Xi_b'\pi$, while $\Xi_b|J^P=\frac{5}{2}^+,2\rangle_{\rho\rho}$ decays dominantly via $\Sigma_b^*K$ and $\Xi_b'^*\pi$. Such moderate states have good potential to be observed in their corresponding dominant decay channels in future experiments.

The three $1D$ $\lambda$-mode $\Xi_b'$ baryons, $\Xi_b'|J^P=\frac{1}{2}^+,1\rangle_{\lambda\lambda}$, $\Xi_b'|J^P=\frac{3}{2}^+,1\rangle_{\lambda\lambda}$ and $\Xi_b'|J^P=\frac{3}{2}^+,2\rangle_{\lambda\lambda}$ are predicted to be moderate states with a width of dozens to a hundred MeV, and mainly decay into the $1P$-wave bottomed baryon via the pionic decay processes. Meanwhile, $\Xi_b'|J^P=\frac{1}{2}^+,1\rangle_{\lambda\lambda}$ and $\Xi_b'|J^P=\frac{3}{2}^+,2\rangle_{\lambda\lambda}$ have significant decay rates into $\Lambda B$, which is likely to be interesting channel for experimental exploration. The other two $1D$ $\lambda$-mode $\Xi_b'$ baryons, $\Xi_b'|J^P=\frac{5}{2}^+,2\rangle_{\lambda\lambda}$ and $\Xi_b'|J^P=\frac{5}{2}^+,3\rangle_{\lambda\lambda}$, have a comparable narrow width of $\Gamma\simeq(15-16)$ MeV, and their decays are dominated by the $\Xi_b'|J^P=5/2^-,2\rangle_{\lambda}\pi$ and $\Xi_b\pi$ channels. $\Xi_b'|J^P=\frac{7}{2}^+,3\rangle_{\lambda\lambda}$ are predicted to be a fairly narrow state with a width of $\Gamma\simeq3$ MeV. The mainly decay channels are $\Lambda_bK$, $\Xi^{'}_{b}\pi$, $\Xi^{'}_{b}\ket{J^P= 3/2^-,1}\pi$ and $\Xi^{'}_{b}\ket{J^P= 5/2^-,2}\pi$.

For the $1D$ $\rho$-mode $\Xi_b'$ baryons, the two states $\Xi_b'|J^P=\frac{5}{2}^+,3\rangle_{\rho\rho}$ and $\Xi_b'|J^P=\frac{7}{2}^+,3\rangle_{\rho\rho}$ have a possibility to be observed in their dominant decay channels $\Lambda_bK$, $\Xi_b\pi$ and $\Xi_b'^*\pi$ for their not broad widths. While, it will be a great challenge to observe the other four states $\Xi_b'|J^P=\frac{1}{2}^+,1\rangle_{\rho\rho}$, $\Xi_b'|J^P=\frac{3}{2}^+,1\rangle_{\rho\rho}$, $\Xi_b'|J^P=\frac{3}{2}^+,2\rangle_{\rho\rho}$ and $\Xi_b'|J^P=\frac{5}{2}^+,2\rangle_{\rho\rho}$ for their very broad width of several hundreds MeV.

\section*{Acknowledgements }

This work is supported by the National Natural Science Foundation of China under Grants No.12005013, No.12175065, No.12235018 and No.11947048.

	\bibliography{Xib}
\end{document}